\documentclass[12pt]{article}
\usepackage{latexsym}
\usepackage[dvips]{epsfig}
\newcommand{\be}{\begin{equation}}
\newcommand{\ee}{\end{equation}}
\newcommand{\lag}{\mathcal{L}}
\newcommand{\act}{\mathcal{A}}
\newcommand{\cs}{S}
\newcommand{\cst}[2]{\cs^{#1}_{#2}}
\newcommand{\detg}{\sqrt{-g}}
\newcommand{\ricci}[2]{R^{#1}_{#2}}
\newcommand{\kro}[2]{\delta^{#1}_{#2}}

\begin{document}

 \rightline{EFI 98-28}

\thispagestyle{empty}
 \begin{center} \large {\bf Born-Regulated Gravity in
Four Dimensions}

 \medskip\normalsize 
James A. Feigenbaum\\

             {\em Enrico Fermi Institute and Department of Physics\\
             The University of Chicago, Chicago, IL 60637, USA}

 \end{center}
 \bigskip
 \bigskip
\bigskip \bigskip \bigskip

\noindent{\bf Abstract}:  Previous work involving Born-regulated gravity
theories in two dimensions is extended to four dimensions.  
The action we consider has {\em two} dimensionful parameters.  Black hole
solutions are studied for typical values of these parameters.  
For masses above a critical value determined in terms of these
parameters, the event horizon persists.  For masses below this
critical value, the event horizon disappears, leaving a ``bare mass'',
though of course no singularity.

\newpage
\thispagestyle{empty}
\mbox{}
\setcounter{page}{0}
\newpage
\bigskip
\section{Introduction}
\bigskip

Recent developments in the theory of strings and branes have
renewed interest in Born-Infeld Lagrangians \cite{BI}
and their non-Abelian generalization \cite {P, L, CM, G, TS}.  In 
the 1930s, Born proposed a modified electromagnetic Lagrangian
which removes the point-charge singularity that mars classical
electrodynamics.  If the Lagrangian is a nonpolynomial function of 
$F_{\mu \nu} F^{\mu \nu}$ with a branch point, then this branch
point can impose an upper bound on field strengths, above which
the Lagrangian will become imaginary.  Specifically Born 
considered the theory \cite{B}
\be
\lag = \Lambda^2 \left[ \sqrt{1 - 
\frac{F_{\mu \nu} F^{\mu \nu}}{2 \Lambda^2}} - 1 \right],
\label{Born}
\ee
which requires $E^2 \leq \Lambda^2$.

Similar theories can be constructed for gravity, replacing the 
Maxwell field tensor with the Riemann curvature tensor.  It is widely
expected that quantum effects remove the singularities
of classical general relativity, cutting off curvatures at the
string scale.  By integrating out all non-gravitational degrees of
freedom in the full Lagrangian for the universe, one can obtain
an effective Lagrangian for gravity which will be nonpolynomial
in curvature components.  This effective Lagrangian might be
of the Born variety, and it is this possibility which we wish to
explore in this paper.

Lagrangians of this type in two dimensions were considered by
Feigenbaum, Freund, and Pigli in Ref. \cite{FFP}, where the
four-dimensional case was briefly alluded to.  Deser and Gibbons
have considered the four-dimensional gravitational analog of
the Born-Infeld Lagrangian \cite{DG}.  For reasons of
simplicity, we will consider here black holes smoothed by an
ordinary Born Lagrangian analogous to Eq.~(\ref{Born}).

In Section Two, we introduce the specific Born-regulated
gravitational Lagrangian which we investigate here.
Remarkably, on account of the two dimensionful parameters in this
Lagrangian, we find two regimes.  In one regime,
an event horizon is present, as in the Einstein-Hilbert case, even though
there is no singularity to ``protect''.  In the other
regime, there is no singularity {\em and} no event horizon.  One
has a ``bare mass'', the regularized version of a naked singularity.
In Sections Three and Four we present an example of both kinds of 
solution.  Then, in Section Five, we will explore the regions in 
parameter space where each of these two types of black hole 
solution occur.

\section{A Lagrangian in Four Dimensions}

In Ref. \cite{FFP}, Born-regulated gravity theories in two dimensions 
were considered.  The action
\be
\act = \int d^2x \detg R [\ln R + \beta \ln(a - R)]
\ee
has Witten black hole solutions \cite{W, CGHS, WA} in the limit
as $\beta \rightarrow 0$ but imposes the bound $R < a$ on the
scalar curvature for $\beta \neq 0$.  As a result, for $\beta \neq 0$,
instead of becoming singular the space-time goes asymptotically
into a de Sitter space with $R = a$.  Note that in two
dimensions the scalar curvature is the sole independent curvature
component.  

In generalizing the notion 
of Born-regulated gravity theories to four dimensions, we must 
recognize
that we now have twenty independent curvature components to play
with and three scalar invariants which can be formed from the
Riemann tensor and which can appear in a Lagrangian:  i.e.
$R_{\mu \nu \rho \sigma} R^{\mu \nu \rho \sigma}$, 
$R_{\mu \nu} R^{\mu \nu}$, and $R$.  We also have empirical data
to contend with in four dimensions, so preferably a gravitational
Lagrangian should reduce to the Einstein-Hilbert Lagrangian in
the weak-field limit.

A candidate action to consider is
\begin{eqnarray}
\act & = & \int d^4x \detg \nonumber \\
& \times & [R + \beta (\sqrt{1 - k_1 \cs - k_2 R_{\mu \nu} R^{\mu \nu}
- k_3 R^2} - 1)],
\label{iniact}
\end{eqnarray}
where $\cst{\alpha}{\beta} \equiv R^{\alpha \mu \nu \rho} 
R_{\beta \mu \nu \rho}$ and $\cs$ is the trace of this tensor.
For the Schwarzschild black-hole solution, $R_{\mu \nu} = 0$.  
Assuming that $R_{\mu \nu} \sim 0$ for black hole solutions to 
the field equations obtained by varying this action, we simplify the 
action by setting $k_2 = k_3 = 0$
and $k_1 = k$ to obtain the action
\be
\act = \int d^4x \detg [R + \beta (\sqrt{1 - k \cs} - 1)],
\label{action}
\ee
which imposes the bound $\cs \leq \frac{1}{k}$ on the square
of the Riemann tensor for $\beta \neq 0$.

The action Eq.~(\ref{action}) yields the field equations
\be
\ricci{\alpha}{\beta} - \frac{1}{2} \kro{\alpha}{\beta} 
[R + \beta (V - 1)] - \frac{k \beta \cst{\alpha}{\beta}}{V}
- 2 k \beta \nabla^{\mu} \nabla_{\nu} 
\left( \frac{R_{\beta \mu}^{\alpha \nu}}{V} \right) = 0
\label{fieldeq},
\ee
where 
\be
V = \sqrt{1 - k \cs}.
\label{V}
\ee

Two parameters, $\beta$ and $k$ appear in the action (\ref{action}).
$\beta$ has dimension $({\rm length})^{-2}$, and the dimension of
$k$ is $({\rm length})^4$.  There will thus be {\em two} scales in
the problem, not unlike string theory.  This, as we shall see in
Section 4, will have as an important consequence the
existence of a critical mass below which the regulated analog to
the black hole solution sheds its event horizon.

\section{Black Hole Solutions for Small $k$ and $\beta$}

We wish to consider solutions which behave as black holes at large
distances and satisfy a spherically symmetric Ansatz for the
metric:
\be
ds^2 = -f^2(r) dt^2 + \frac{dr^2}{h^2(r)} + r^2(d\theta^2 + 
\sin^2(\theta) d\varphi^2).
\label{schmet}
\ee
Inserting this Ansatz into Eqs.~(\ref{fieldeq}), we obtain three
nontrivial equations corresponding to the variation
of the action with respect to $g_{tt}$, $g_{rr}$, and 
$g_{\theta\theta}$ (the equations corresponding to $g_{\theta\theta}$
and $g_{\varphi\varphi}$ being identical).  However since there are 
only two unknown functions in the Ansatz, $f(r)$ and $h(r)$, these three 
equations are not independent.

In the Schwarzschild solution for a black hole of mass $M$,
\be
f_s(r)=h_s(r)=\sqrt{1 - \frac{2M}{r}}.
\ee
Consequently in the limit of small $k$ and $\beta$, for $r \rightarrow \infty$
we write
\be
f(r) = \sqrt{1 - \frac{2M}{r}}(1 + \phi(r))
\label{eqf}
\ee
and
\be
h(r) = \sqrt{1 - \frac{2M}{r}}(1 + \eta(r)),
\label{eqh}
\ee
where $\eta(\infty) = \phi(\infty) = 0$.
Let
\be
\lambda = \frac{k^2 \beta}{(2 M)^6}.
\label{lambda}
\ee
This dimensionless parameter
characterizes perturbations of the Schwarzschild solution.
Solving Eqs.~(\ref{fieldeq}) for $\phi$ and $\eta$ to lowest order in
$\lambda$, we find
\be
\phi(r) = \frac{-8 k^2 M^3 \beta}{r^9} 
\left(\frac{8 r - 11 M}{r - 2 M}\right) 
+ O \left( \frac{k^3 \beta^2 M^3}{r^{11}} \right)
\label{s3}
\ee
and
\be
\eta(r) = \frac{-8 k^2 M^3 \beta}{r^9} 
\left(\frac{36 r - 67 M}{r - 2 M}\right) 
+ O \left( \frac{k^3 \beta^2 M^3}{r^{11}} \right).
\label{s4}
\ee
Clearly in the limit of small $k$ and $\beta$, deviations from
the Schwarzschild solution are negligible for $r \gg 2M$.  

We may note that the $\frac{k^2 M^3 \beta}{r^9}$ dependence of the
prefactors in $\phi(r)$ and $\eta(r)$ is easily understood.  For
the unperturbed Schwarzschild solution, the lowest order nonvanishing
terms in the field equations (\ref{fieldeq}) derive from the $\beta k^2 S^2$
term in the expansion of the Born-regulating square root of the 
Lagrangian (\ref{action}), and to lowest order in $\frac{1}{r}$ these 
terms go as $\frac{\beta k^2 M^4}{r^{12}}$.
By contrast, the inclusion of $\phi(r)$ and $\eta(r)$ in the solution of
Eqs. (\ref{eqf}),(\ref{eqh}) leads to nonvanishing Ricci tensor and
scalar curvature terms in the field equations which go as $\frac{M \phi}{r^3}$
and $\frac{M \eta}{r^3}$ to lowest order in $\frac{1}{r}$.  Consequently,
in order for all these terms to cancel, we must have the prefactors seen
in Eqs. (\ref{s3}),(\ref{s4}).

To analyze these solutions near and within the event horizon
$r \approx 2M$, we must transform to Kruskal-like coordinates,
exchanging $r$ and $t$ for the light-cone coordinates $u$ and $v$.  
The spherically symmetric Ansatz analogous to Eq.~(\ref{schmet}) for
Kruskal-like coordinates is
\be
ds^2 = -\exp(2\rho(w)) du dv + r^2(w) (d\theta^2 + 
\sin^2(\theta) d\varphi^2),
\label{schkru}
\ee
where $w = uv$ and the functions $\rho$ and $r$ are functions
of $w$ alone.  Here $r(w)$ is precisely the coordinate $r$
in the Schwarzschild-like Ansatz of Eq.~(\ref{schmet}).

Inserting this Ansatz into Eqs.~(\ref{fieldeq}), we again obtain
three separate but presumably not independent equations
corresponding to the variation of Eq.~(\ref{action}) with
respect to $g_{uu}$, $g_{uv}$, and $g_{\theta\theta}$.
In order to integrate these differential equations, continuing
from our solution of Eqs.~(\ref{eqf}),(\ref{eqh}), we must know
$r$ and $\rho$ and their first derivatives at some point.

The event horizon in $u$-$v$ coordinates is the surface $w=0$.  
In the region $w < 0$ which corresponds to the region outside
the event horizon, we can make the coordinate transformation
\be
u = - \sqrt{-w(r)} \exp \left(\frac{-t}{4M} \right)
\label{transu}
\ee
and
\be
v = \sqrt{-w(r)} \exp \left(\frac{t}{4M} \right).
\label{transv}
\ee
Here $w(r)$ is the inverse of the function $r(w)$ in Eq.~(\ref{schkru}).  
The corresponding transformation of
the metric then gives
\be
f(r) = \frac{\sqrt{-w(r)} \exp(\rho(w(r)))}{4M}
\ee
and
\be
h(r) = \frac{-2 \sqrt{-w(r)} \exp(-\rho(w(r)))}{w'(r)}.
\ee
We see from this identification that the event horizon in 
Schwarzschild-like coordinates is $r = r_s$, where 
\be
f(r_s) = h(r_s) = 0.
\label{scwrad}
\ee
Solving Eqs.~(\ref{scwrad}) for $r_s$, we find to first order
in $\lambda$ that $r_s = 2M(1 + 5 \lambda)$.  

As always, at the event horizon, there is a coordinate singularity in $r$-$t$ 
coordinates, but the Riemann tensor remains finite.  Using
the coordinate transformation of Eqs.~(\ref{transu}) and (\ref{transv}),
we can relate the components of the Riemann tensor in the two coordinate 
systems:
\be
R^{uv}_{uv} = R^{tr}_{tr},
\label{uvuv}
\ee
\be
R^{u\theta}_{u\theta} = R^{v\theta}_{v\theta} = 
\frac{1}{2} (R^{t\theta}_{t\theta} + R^{r\theta}_{r\theta}),
\label{utut}
\ee
and
\be
\frac{v}{u} R^{u\theta}_{v\theta} = \frac{u}{v} R^{v\theta}_{u\theta}
= \frac{1}{2} (R^{r\theta}_{r\theta} - R^{t\theta}_{t\theta}).
\label{utvt}
\ee

We expand $r(w)$ and $\rho(w)$ around $w=0$:
\be
r(w) = r_s (1 + \sum_{n=1}^{\infty} c_n w^n),
\label{eqr}
\ee
and
\be
\rho(w) = -\frac{1}{2} \ln(A) + \sum_{n=1}^{\infty} a_n w^n.
\label{eqrho}
\ee
The scale of $w$ is arbitrary, so we can set $c_1 = \exp(-1)$, which
for the Schwarzschild solution would place the curvature singularity
at $w=1$.  Using Eqs.~(\ref{uvuv})-(\ref{utvt}), we compare at the event
horizon the Riemann tensor components in the
solution of Eqs.~(\ref{eqf})-(\ref{s4}), written in Schwarzschild-like
coordinates, with the series expansion of Eqs.~(\ref{eqr}),(\ref{eqrho}),
written in Kruskal-like coordinates.  In this manner, 
we obtain values for $A$, $a_1$, and $c_2$.  This
is the remaining information necessary to integrate Eqs.~(\ref{fieldeq}).

As an example of a numerical solution, we choose $k = 1$, $M = 100$, and 
$\beta = 10^9$ to obtain a small value of the perturbation parameter 
$\lambda = .000015625$ from Eq.~(\ref{lambda}).
In Fig.~1, $\cs$ is plotted as a function of $w$ for the Schwarzschild
solution and the perturbed solution calculated to $O(w^{40})$.  We see
that as $w \rightarrow 1$, where the Schwarzschild solution is singular,
$\cs$ for the perturbed solution is less than $\cs$ for the Schwarzschild
solution.  However to fortieth order in $w$, $\cs$ remains much less than 
the upper bound of $\cs \leq 1$ at $w=1$.  In order to see the upper
bound come into effect, we would need to calculate the solution to
a very high order with such small values of $k$ and $\beta$.
In the next section, we will consider a solution with much larger
values of these parameters, where the curvature bound will become
evident.

\begin{figure}
\vspace{-1 in}
\epsfig{file=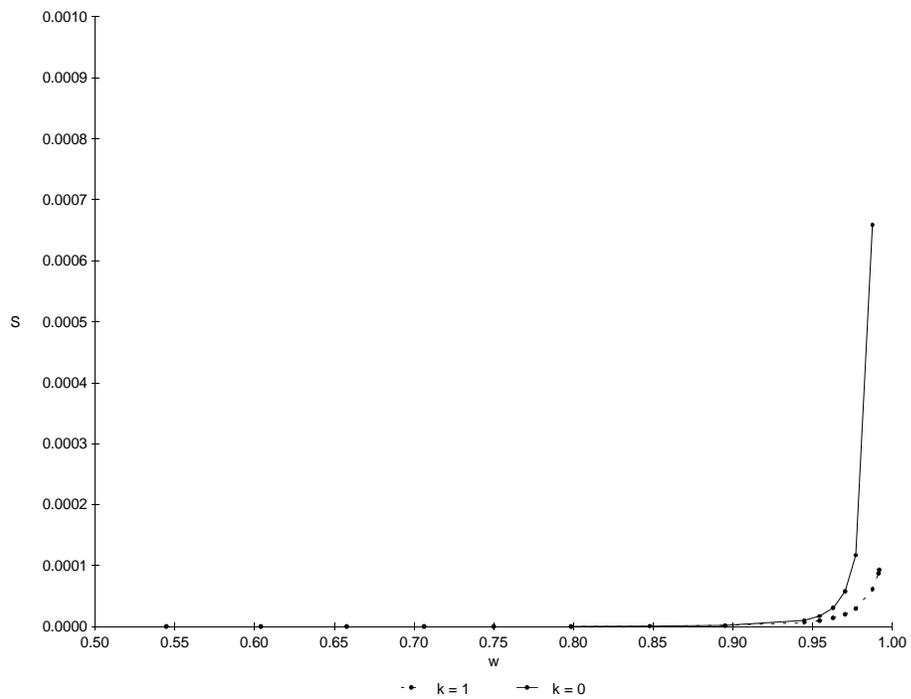, width =4.5 in, height = 6 in, angle = -90} 
\caption{The curvature invariant                                  
$S = R_{\alpha \beta \gamma \delta} R^{\alpha \beta \gamma \delta}$      
as a function of the Kruskal-coordinate combination $w = uv$ for         
the ordinary Schwarzschild solution (full curve) and for the             
Born regulated theory (dashed curve) to order $w^{40}$ 
with $k=1$ and $\beta = 10^9$, 
both for an object of mass 100.}                     
\end{figure}                                                             

\section{Black Hole Solutions for Large $\beta$}

For sufficiently large values of $\beta$, we can
ignore the scalar curvature term of Eq.~(\ref{action}) and 
the field equations reduce to 
\be
\frac{1}{2} \kro{\alpha}{\beta} (V - 1) 
+ \frac{k \cst{\alpha}{\beta}}{V} + 2 k \nabla^{\mu} \nabla_{\nu} 
\left( \frac{R_{\beta \mu}^{\alpha \nu}}{V} \right) = 0.
\label{Sfieldeq}
\ee
Again, we wish to find solutions of these field equations that act
like a black hole solution of mass $M$ as $r \rightarrow \infty$.  
In order to expand around infinity, we introduce the variable
$q = \frac{1}{r}$.  Replacing $r$ with $q$ in Eq.~(\ref{schmet})
we have the equivalent Ansatz:
\be
ds^2 = -f^2(q) dt^2 + \frac{dq^2}{q^4 h^2(q)} 
+ \frac{d^2\theta + \sin^2(\theta) d\varphi^2}{q^2}.
\label{qmet}
\ee
We expand $f(q)$ and $h(q)$ around $q=0$:
\be
f(q) = 1 + \sum_{n=1}^{\infty} b_n q^n
\ee
and
\be
h(q) = 1 + \sum_{n=1}^{\infty} d_n q^n.
\ee
We require $f$ and $h$ to satisfy the same boundary conditions as
the Schwarzschild solution, so we set $b_1 = d_1 = -M$.  Then,
if we solve recursively for the higher order coefficients, we obtain
\be
f(q) = \sqrt{1 - 2 M q} + \frac{256 k M^3 q^7}{336} + O(q^8)
\ee
and
\be
h(q) = \sqrt{1 - 2 M q} + \frac{128 k M^3 q^7}{48} + O(q^8).
\ee
Thus in the infinite $\beta$ limit, this metric is indistinguishable
from the Schwarzschild metric far from the black hole.

\begin{figure}
\vspace{-1 in}
\epsfig{file=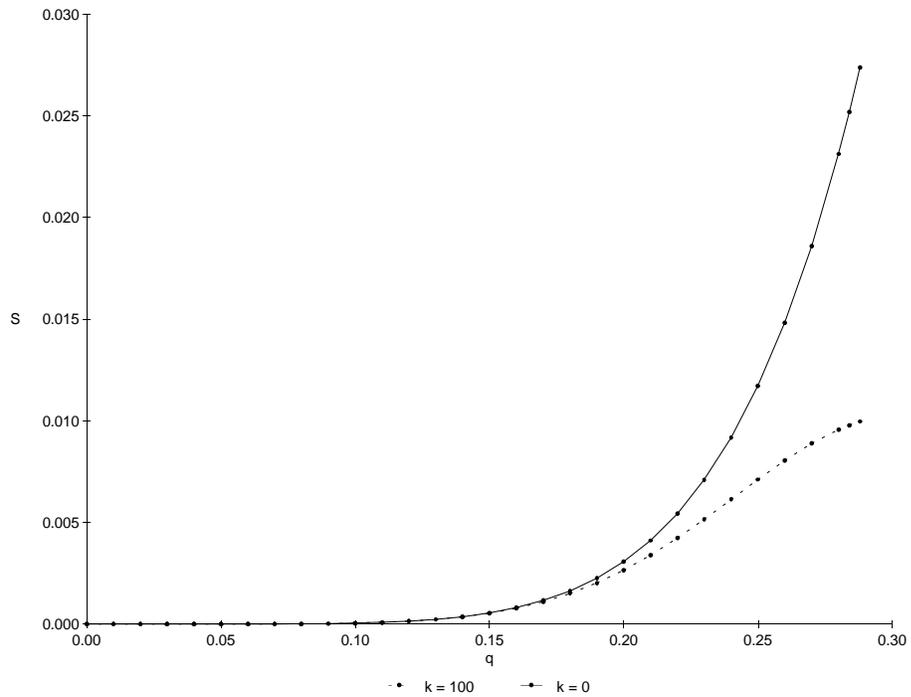, width = 4.5 in, height = 6 in, angle = -90}
\caption{The curvature invariant 
$S = R_{\alpha \beta \gamma \delta} R^{\alpha \beta \gamma \delta}$
as a function of $q = \frac{1}{r}$ for
the ordinary Schwarzschild solution (full curve) and for the
Born regulated theory (dashed curve) with $k = 100$, both for an object
of unit mass.}
\end{figure}

In Fig. 2, we plot $\cs$ versus $r$ for our solution here with $M=1$ and 
$k=100$ along with the Schwarzschild solution for $M=1$.  Here the
curvature bound $\cs \leq .01$ is quite evident.  As $q \sim .29$,
$\cs$ approaches .01.  Numerical integration past $q = .29$ becomes
exceedingly difficult but is fortunately unnecessary.
Indeed, as $q \rightarrow \infty$, one can infer from the 
action principle that the solution goes 
asymptotically into
a solution of constant $\cs$.  With $V$ as defined in Eq.~(\ref{V}), the 
field equations, Eqs~(\ref{fieldeq}), can be 
rewritten as
\begin{eqnarray}
\frac{1}{2} V^5 (1 - V) \kro{\beta}{\alpha} & = & \nonumber \\
& & k V^4 [\cst{\beta}{\alpha} 
+ 2 \nabla^{\mu}\nabla_{\nu} R_{\alpha \mu}^{\beta \nu}] \nonumber \\
& + & k^2 V^2 [(\nabla^{\mu} R_{\alpha \mu}^{\beta \nu})(\nabla_{\nu} \cs)
+ (\nabla_{\nu} R_{\alpha \mu}^{\beta \nu})(\nabla^{\mu} \cs)
+ R_{\alpha \mu}^{\beta \nu}(\nabla^{\mu} \nabla_{\nu} \cs)] \nonumber \\
& + & \frac{3}{2} k^3 R_{\alpha \mu}^{\beta \nu} (\nabla^{\mu} \cs)
(\nabla_{\nu} \cs).
\end{eqnarray}
If $\cs$ is constant at $\frac{1}{k}$, then $V = 0$, and clearly the field
equations are satisfied.  So if we have $\cs \rightarrow \frac{1}{k}$
on the surface $q = q_0$, it follows that $\cs = \frac{1}{k}$ for $q > q_0$.

Note that one glaring absence from the solution with $M=1$ and $k=100$
that we have described here is a coordinate singularity.  In fact
there can be no coordinate singularity for finite $q$.  The curvature 
components in
$q$-$t$ coordinates are
\be
R^{tq}_{tq} = -\frac{q^2h}{f}(q^2 h f')',
\ee
\be
R^{t\theta}_{t\theta} = q^3 h^2 (\ln(f))',
\label{thth}
\ee
\be
R^{q\theta}_{q\theta} = q^3 h h',
\ee
and
\be
R^{\theta\varphi}_{\theta\varphi} = q^2 (1 - h^2).
\label{hphp}
\ee
Since $\cs = 4(R^{tq}_{tq})^2 + 8(R^{t\theta}_{t\theta})^2
+ 8(R^{q\theta}_{q\theta})^2 + 4(R^{\theta\varphi}_{\theta\varphi})^2$,
we have the constraint 
\be
R^{\theta\varphi}_{\theta\varphi} \leq \frac{1}{2 \sqrt{k}}.
\label{constraint}
\ee
For $q \geq .29$ and $k=100$, it follows from Eq.~(\ref{hphp}) and
the constraint (\ref{constraint}) that 
$.6368 \leq h \leq 1.2627$.  Since $h$ and $R^{t\theta}_{t\theta}$
must be finite, it also follows, from Eq.~(\ref{thth}), that 
$\frac{d}{dq}(\ln(f))$ must be finite, and so $\ln(f)$
and $f$ must remain finite for finite $q$.  

Since $f$ and $h$ must remain finite for finite $q$, it follows that
there can be no coordinate singularity and therefore no event horizon
for finite $q$.  The solution we have here describes what is not 
really a black hole but a ``bare mass".  It is important to note that this
is {\em not} a ``naked singularity".  Although it is ``bare'' or ``naked'' 
in the sense that it is not hidden behind an event horizon, it is 
not a ``naked singularity'' because there is no singularity.

\section{Shedding the Event Horizon}

The absence of an event horizon is not a universal property
of all solutions to Eqs.~(\ref{fieldeq}) which behave as black
holes for large $r$.  If there is an 
event horizon, we must have $h=0$ at this horizon.  Then (\ref{hphp}) and
(\ref{constraint}) imply that the reciprocal Schwarzschild radius $q_s$ 
satisfies $q_s^2 \leq \frac{1}{2 \sqrt{k}}$.  
For small $\beta$, $q_s \sim \frac{1}{2M}$, so it follows that 
the dimensionless ratio 
$\frac{k}{M^4}$ will determine whether there 
can be an event horizon.  
For $\frac{k}{M^4} \gg 1$, the event horizon must disappear.  
For $\frac{k}{M^4} \ll 1$, there should still
be an event horizon as in Section 3.  Since we used the original, unsimplified
field equations (\ref{fieldeq}) in Section 3, 
it is important here to recognize that the argument at the end of Section
4 depends on the metric Ansatz (\ref{qmet}) and the curvature bound 
$S \leq \frac{1}{k}$ but not on the details of the Born Lagrangian.  As
such, this argument applies equally well for small $\beta$,
the case considered here, as for large $\beta$, the case covered in
Section 4.
For a given value of the parameter $k$ appearing in the Lagrangian,
the event horizon will disappear as the mass falls below some critical
mass of order $k^{1/4}$ (or $\frac{k^{1/4}}{G}$ if we include Newton's
gravitational constant explicitly).  

For very small (yet nonzero!) $\beta$, what makes this mechanism feasible
is the presence of two dimensionless parameters in
the problem, $M^2 \beta$ and $\frac{k}{M^4}$, corresponding to the 
two dimensionful parameters $k$ and
$\beta$ in the action.  We can choose $\beta M^2$ to be
arbitrarily small.  However, if we also choose $\frac{k}{M^4} \gg 1$, then 
at the place where one would naively expect an event horizon to 
appear, $k S \rightarrow 1$.
As a result, even though each individual term in the series 
expansion of $\beta \sqrt{1 - k S}$ in the action (\ref{action}) 
may be very small, the terms do not
diminish in magnitude.  Consequently, their infinite sum will still
dominate over the Einstein-Hilbert term, giving rise to a very different
solution from ordinary Schwarzschild.  Besides being nonsingular, this
solution lacks an event horizon. 

The precise value of the critical mass will depend on the value of $\beta$.  
The surface in $k$-$\beta$-$M$ space where the transition between having
and not having an event horizon occurs should evidently take the form
\be
\frac{k}{M^4} = \sigma(M^2 \beta),
\ee
where $\sigma$ is an unknown function.
For general values of $M^2 \beta$, the precise form of 
$\sigma(M^2 \beta)$ will have to be determined numerically in
the manner of the last section.  
In this way, we have found an upper bound on $\sigma$ for large
$\beta$ of $\sigma(M^2 \beta) < 1$.

One can infer the limiting value of $\sigma(M^2 \beta)$ as 
$\beta \rightarrow 0$.  For very small values of $\beta$, solutions
of Eqs. (\ref{fieldeq}) with mass $M$ should behave exactly like the
Schwarzschild solution with mass $M$ up until the point where 
$k S \rightarrow 1$, at which point $S$ will flatten out and asymptotically
approach the value $\frac{1}{k}$.  For Schwarzschild, $S = \frac{3}{4 M^4}$
at the event horizon.  So if $\frac{3}{4 M^4} < \frac{1}{k}$, there
should still be an event horizon since the solution will not begin to 
deviate from Schwarzschild until we are inside the black hole.  If
$\frac{3}{4 M^4} > \frac{1}{k}$, the solution will deviate from Schwarzschild
before an event horizon can occur, and so there can be no event horizon.  
Thus we conclude that
\be
\lim_{\beta \rightarrow 0} \sigma(M^2 \beta) = \frac{4}{3}.
\ee

Evidently $\sigma(M^2 \beta)$ decreases with $\beta$ and so
the critical mass, below which the event horizon disappears, 
increases with $\beta$, as one would intuitively expect since it would
be very surprising if the effects of Born regulation should become less
apparent as we increase $\beta$.

\section{Conclusion}

In this paper, we have investigated a Born-regulated theory of
gravity in four dimensions.  We have found that solutions to this
theory exist which behave asymptotically as black holes of mass $M$
but become spaces of constant 
$\cs=R_{\alpha \beta \gamma \delta} R^{\alpha \beta \gamma \delta}$ 
at small radii.  These spaces of
constant $\cs$ are analogous to the de Sitter spaces which we found in 
\cite{FFP} and which Brandenberger found in \cite{B1, B2}.

For large values of $\frac{k}{M^4}$ in the Born-regulated
Lagrangian, there is no event horizon in these solutions, and we
have a ``bare mass'' instead of a black hole.  If we assume that $k$ is 
valued at the Planck scale, then the event horizon will be absent
only for black hole masses below the Planck scale.

\section{Acknowledgements}
We would like to thank Peter Freund and Mircea Pigli for their prior 
collaboration
and advice on this topic.  We would also like to thank the members of
the thesis committee, Susan Coppersmith, David Kutasov, Gene Mazenko,
and Frank Merritt, for 
some helpful insights.  This work was 
supported in part by NSF Grant No. PHY-9123780-A3.

\end{document}